\documentclass{article}
\usepackage{spconf,amsmath,graphicx}
\usepackage{multirow}
\usepackage{bm}
\usepackage{array}
\usepackage{booktabs}
\usepackage{adjustbox}
\usepackage[marginal]{footmisc}
\usepackage{marvosym}
\title{Unit selection synthesis based data augmentation for fixed phrase speaker verification}
\name{Houjun Huang$^{1}$$^{,2}$, Xu Xiang$^{1}$, Fei Zhao$^1$, Shuai Wang$^{2}$, \textsuperscript{\Letter}Yanmin Qian$^{2}$\thanks{Yanmin Qian is the corresponding author. This work was supported by the National Key R\&D Program of China (No. 2018YFB1004602) and the China NSFC project (No. 62071288).}}

\address{
$^{1}$ AISpeech Ltd, Suzhou China\\
$^{2}$MoE Key Lab of Artificial Intelligence, AI Institute SpeechLab, Department of Computer Science\\and Engineering Shanghai Jiao Tong University, Shanghai,China \\
\{houjun.huang, xu.xiang, fei.zhao01\}@aispeech.com, \{feixiang121976,yanminqian\}@sjtu.edu.cn}
\newcommand{\etal}{\textit{et al.}}
\begin{document}
%\ninept
%
\maketitle
\begin{abstract}
Data augmentation is commonly used to help build a robust speaker verification system, especially in limited-resource case. However, conventional data augmentation methods usually focus on the diversity of acoustic environment, leaving the lexicon variation neglected. For text dependent speaker verification tasks, it's well-known that preparing training data with the target transcript is the most effectual approach to build a well-performing system, however collecting such data is time-consuming and expensive. In this work, we propose a unit selection synthesis based data augmentation method to leverage the abundant text-independent data resources. In this approach text-independent speeches of each speaker are firstly broke up to speech segments each contains one phone unit. Then segments that contain phonetics in the target transcript are selected to produce a speech with the target transcript by concatenating them in turn. Experiments are carried out on the AISHELL Speaker Verification Challenge 2019 database, the results and analysis shows that our proposed method can boost the system performance significantly.

\end{abstract}
\begin{keywords}
speaker verification, data augmentation, unit selection synthesis, x-vector
\end{keywords}
\section{Introduction}
\label{sec:intro}
Speaker verification (SV) aims to confirm the claimed identity given his/her speech. Considering the constraint on the speech content, the SV task can be further categorized into two classes: text-dependent and text-independent. The former requires the same content for the enrollment and test utterances, while the latter doesn't.

Traditional speaker recognition systems are based on statistical models such as Gaussian Mixture Model-Universal Background Model (GMM-UBM)~\cite{reynolds2000speaker}. The i-vector~\cite{dehak2010front} system projects the GMM super-vector to a lower-dimensional and more speaker-discriminative vector. Recently, utterance-level deep speaker embedding methods such as x-vector~\cite{snyder2017deep,snyder2018x}, have shown better performance than i-vector on many standard speaker recognition data-sets.

Although the SV performance has been improved greatly over the recent years, further upgrading its robustness in real applications is still a challenge task due to the complex environment. Data augmentation is conventionally adopted in building a SV system to improve its robustness. Snyder~\etal in~\cite{snyder2017deep,snyder2018x} manually employed additive noises and reverberation to original speech segments in the training set to train a robust embedding extractor and then extract "clean" and "noisy" embeddings to train probabilistic linear discriminant analysis (PLDA) for both i-vector and x-vector. SpecAugment~\cite{park2019specaugment,wang2020investigation} also shows promising results on the speaker verification task. On the other hand, researchers also apply deep generative models such as generative adversarial network (GAN) and  variational autoencoder (VAE) to generate x-vector embeddings directly to train a robust PLDA~\cite{yang2018generative,wu2019data}. 

Despite the effectiveness exhibited by the data augmentation methods above, they only consider the variation of acoustic environment. Text variation should also be considered to build a robust SV system, especially for text-dependent tasks. In practice, to achieve the state-of-the-art performance, text-dependent SV requires the same set of text to be spoken during the training stage and the test stage.~\cite{yang2020text,qin2019far,qin2020hi}. When there is no or limited training data with the designated phrase to build a text-dependent SV system, those data augmentation approaches couldn't help to get promising performance. To improve the system performance from this aspect, in this work, we propose a novel method which generates new speech containing designated phrase from text-independent database using the \emph{unit selection synthesis}~\cite{hunt1996unit,black1997automatically,conkie1999robust}.

\section{Unit selection synthesis based data augmentation}
In the case that limited or no text-dependent training data is available to build a text-dependent SV system, generating speech utterances with the fixed phases of more speakers is the most effectual solution. If a text-independent database with a large number of units is available, speaker discriminative synthesized speeches can be produced by concatenating the wave-forms of units selected from speeches of each speaker~\cite{hunt1996unit,black1997automatically,conkie1999robust}. 

\begin{figure}[htb]
\centering{\includegraphics[width=80mm]{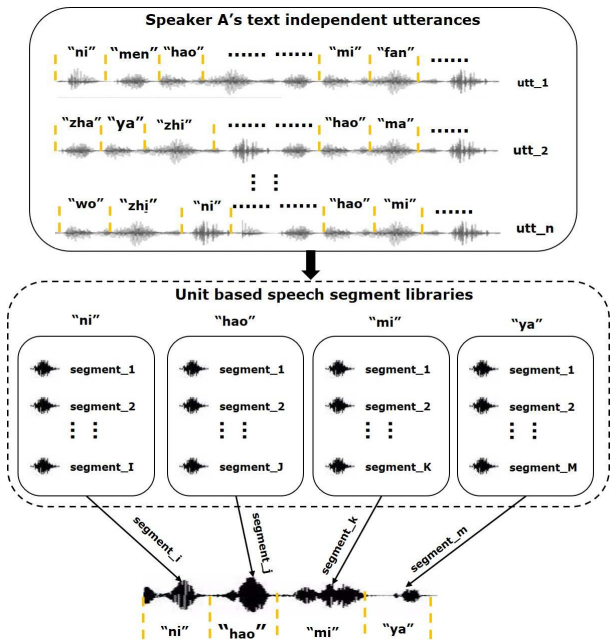}}
\caption{Generate ``ni hao, mi ya'' using unit selection synthesis for a speaker in the text-independent database}
\label{fig:make-up}
\end{figure}

% In Figure~\ref{fig:make-up}, 
Here, we use the Chinese wake-up word "ni hao mi ya" as an example to describe how to carry out the proposed approach. In this work, we treat each Chinese character as the phonetic unit. Thus, "ni hao mi ya" can be converted to a phonetic unit sequence "ni"-``hao"-``mi"-``ya." The unit selection synthesis based augmentation is shown in Figure~\ref{fig:make-up}, which contains the following steps,
\begin{enumerate}
    \item For each speaker in the text-independent database, we chunk his speech into segments which only contain single characters.
    \item Select all the segments which contain desired phonetic units to generate phonetic unit libraries.
    \item For each speaker, synthesize new speech containing the target transcript by concatenating the sampled units from the phonetic unit libraries.
\end{enumerate}

The standard unit selection synthesis system aims to produce more natural-sounding synthesized speeches, while the proposed approach aims at generating utterances with fix phonetic content that are speaker discriminative and variable to train more robust text-dependent SV systems. Speech segments are randomly selected from the same speaker for each unit and no flexible join technique is used in concatenating to ensure diversity of synthesized data. In our experiments, N$_{i}$ utterances will be synthesized for speaker \textit{i} where N$_{i}$ is the max size of his/her unit based speech segment libraries.

\section{Experiments}
\label{sec:experiments}
\subsection{Experimental data}
In this work, we use the AISHELL2\footnote{AISHELL2 is publicly available at http://www.aishelltech.com/aishell\_2.}~\cite{du2018aishell} as text-independent training data, AISHELL-wakeup\footnote{AISHELL-wakeup is available at http://openslr.org/85/.}~\cite{qin2020hi} as text-dependent training data, and AISHELL-2019B-eval dataset\footnote{Speech data of AISHELL-2019B-eval and trial files are available at http://openslr.org/85/.}~\cite{qin2020hi} as text-dependent test set.

The AISHELL2 is a 1000-hour Mandarin Chinese Speech Corpus that contains 1003664 close-talk utterances from 1991 speakers. The speech utterance contains several domains, including keywords, voice command, smart home, autonomous driving, industrial production, etc. The recording was put in a quiet room environment using an iOS system mobile phone.

The AISHELL-wakeup database has 3,936,003 utterances from 254 speakers. The content of utterances covers two wake-up words: "ni hao, mi ya" in Chinese and "Hi, Mia" in English. The recording process was put in a real smart home environment where one close-talking microphone was placed 25cm away from the speaker, six 16-channel circular microphone arrays were placed around the person with a distance including 1m, 3m and 5m from the speaker and the noise source was randomly placed close to one of the microphone arrays. The 993083 mono channel wave-files of the Chinese wake-up word "ni hao, mi ya" are chosen to train the text-dependent model in our experiments. 

The AISHELL-2019B-eval contains recordings of 86 speakers with Chinese wake-up word "ni hao, mi ya".The room setting and recording devices are the same as that of AISHELL-wakeup. Utterances of the last 44 people are selected as the test set since they are more challenging~\cite{qin2020hi}. This corpus has two tasks: close-talking enrollment task (utterances from the close-talking mic are used for enrollment) and far-field enrollment task (utterances from one 16-channel circular microphone array which is 1m away from the speaker are used for enrollment). The testing data for both tasks are far-field utterances recorded with 16-channel circular microphone arrays. 

\subsection{Experimental setups}
The frame-level alignments of the above databases are generated using the official Kaldi~\cite{povey2011kaldi} AISHELL2 speech recognition recipe(s5)~\cite{du2018aishell}, and voice activity detection (VAD) labels of them are generated based on their alignments.

347706 utterances of 1986 speakers are generated from the AISHELL2 database using the proposed approach, and we call this data-set as "AISHELL2-aug" in the following sections. After the data augmentation method proposed in~\cite{snyder2017deep,snyder2018x} is adopted, the number of speech utterances of AISHELL2, AISHELL-wakeup and AISHELL2-aug are extended to 4013020, 3972332 and 3824766. 40-dimensional fbank features are extracted from these databases with a frame shift of 10ms and a window width of 25ms. VAD is employed to filter out non-speech frames. Mean normalization is then applied over a sliding window of up to 300 frames. Finally, SpecAugment~\cite{park2019specaugment} is applied to the fbank features.

The x-vector system~\cite{snyder2017deep,snyder2018x} is used in our experiments. The architecture of the speaker-discriminative TDNN is illustrated in Table~\ref{tab:Architecture-x-vector}. The T in the stats-pool layer corresponds to the frame number of the input features. X-vector embeddings are extracted at layer segment6, before the projection layer. The N in the projection layer corresponds to the number of training speakers. Additive angular margin loss~\cite{deng2019arcface,xiang2019margin} with m=0.2 and s=32.0 is used as the projection layer since it has shown better performance than other losses. For the enrollment or test utterances that have 16 channels from a 6-channel circular microphone array, we adopt the strategy of speaker embedding level averaging as do in~\cite{qin2020hi}.

\begin{table}[h]
    \caption{Architecture of the x-vector}
    \label{tab:Architecture-x-vector}
    \centering
    \begin{tabular}{c|c|c|c}\toprule
        Layer & Layer context & Total context & Input*output\\\midrule
        frame1 & [t-2, t+2] & 5 & 200*256\\\midrule
        frame2 & {t-2, t, t+2} & 9 & 768*256\\\midrule
        frame3 & {t-3, t, t+3} & 15 & 768*256\\\midrule
        frame4 & {t} & 15 & 256*256\\\midrule
        frame5 & {t} & 15 & 256*512\\\midrule
        stats-pool & [0,T) & T & 512T*1024\\\midrule
        segment6 & {0} & T & 1024*256\\\midrule
        projection & {0} & T & 256*N\\\bottomrule
    \end{tabular}
\end{table}

\begin{figure}[htb]
\centering{\includegraphics[width=45mm]{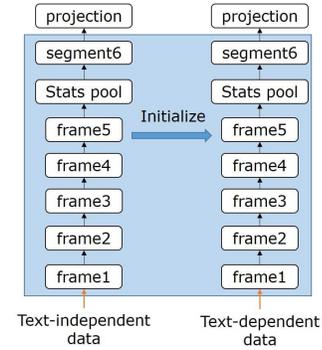}}
\caption{Initialize text-dependent x-vector model with text-independent model}
\label{fig:Initialize}
\end{figure}

AISHELL2 is firstly used to train the text-independent x-vector model. As is shown in Figure~\ref{fig:Initialize},  when we train a text-dependent model, the parameters of layers before the projection layer are initialized by the text-independent x-vector model. To test the effectiveness of the proposed approach, three text-dependent x-vector models are trained with AISHELL-wakeup, AISHELL2-aug and AISHELL-wakeup + AISHELL2-aug, respectively. 

%Initialize text-dependent x-vector model with text-independent model

Cosine similarity serves as back-end scoring method during testing. We report results in Equal error rate (EER), minimum detection cost for P(tar) = 0.01 (minDCF).

\subsection{Experimental results}
\label{sec:results}

Experiment results on close-talking and far-field enrollment task are shown in Table~\ref{tab:close-talking} and Table~\ref{tab:far-field} respectively. 

\begin{table}[!th]
    \caption{Performance on close-talking enrollment task of x-vector models trained with different data. \textit{AISHELL2-aug} is the training data produced by the proposed approach.}
    % Conventional data augmentation methods have been adopted on AISHELL2, AISHELL-wakeup and AISHELL2-aug.
    \label{tab:close-talking}
    \centering
    % \begin{tabular}{m{0.15\textwidth}|c|c}\toprule
    \begin{adjustbox}{width=0.5\textwidth,center}
    \begin{tabular}{c|c|c}\toprule
        \centering train data set & EER(\%) & minDCF\\\midrule
        AISHELL2 & 6.978 & 0.616\\\midrule
        AISHELL-wakeup & 5.796 & 0.606\\\midrule
        AISHELL-wakeup+AISHELL2 & 4.386 & 0.446\\\midrule
        AISHELL2-aug & 4.087 & 0.405 \\\midrule
        AISHELL-wakeup+AISHELL2-aug & \textbf{3.019} & \textbf{0.282}\\\bottomrule
    \end{tabular}
    \end{adjustbox}
\end{table}

\begin{table}[!th]
    \caption{Performance on far-field enrollment task of x-vector modles trained with different data. \textit{AISHELL2-aug} is the training data produced by the proposed approach.}
    % Conventional data augmentation methods have been adopted on AISHELL2, AISHELL-wakeup and AISHELL2-aug.
    \label{tab:far-field}
    \centering
    % \begin{tabular}{|m{0.15\textwidth}|c|c|}\hline
    \begin{adjustbox}{width=0.5\textwidth,center}
    \begin{tabular}{c|c|c}\toprule
        \centering train data set & EER(\%) & minDCF\\\midrule
        AISHELL2 & 5.993 & 0.466\\\midrule
        AISHELL-wakeup & 4.598 & 0.409\\\midrule
        AISHELL-wakeup+AISHELL2 & 3.274 & 0.335\\\midrule
        AISHELL2-aug & 3.673 & 0.319 \\\midrule
        AISHELL-wakeup+AISHELL2-aug & \textbf{2.562} & \textbf{0.224} \\\bottomrule
    \end{tabular}
    \end{adjustbox}
\end{table}

Compared to the text-independent x-vector model, when there is no text-dependent resource to build the SV system, AISHELL2-aug generated by the proposed method trained text-dependent x-vector model achieves a relative performance improvement of 41.4\% in EER and 38.7\% in EER on close-talking enrollment task and far-field enrollment task respectively. Compared to the AISHELL-wakeup+AISHELL2 trained text-dependent x-vector model, when there is limited text-dependent resource to build the SV system, AISHELL-wakeup+AISHELL2-aug trained model obtains further relative performance improvement of 31.16\% in EER and 21.74\% in EER on close-talking enrollment task and far-field enrollment task, respectively. 

\subsection{Analysis}
The mismatch of speech content between training and test data will induce a severe degradation of SV performance. The proposed method aims to produce training speech utterances whose contents match the text-dependent test set. This approach is effective only if the synthesized speeches contain a phonetic context of "ni"-``hao"-``mi"-``ya" and be speaker discriminative.

\begin{figure}[!htb]
\centering{\includegraphics[width=80mm]{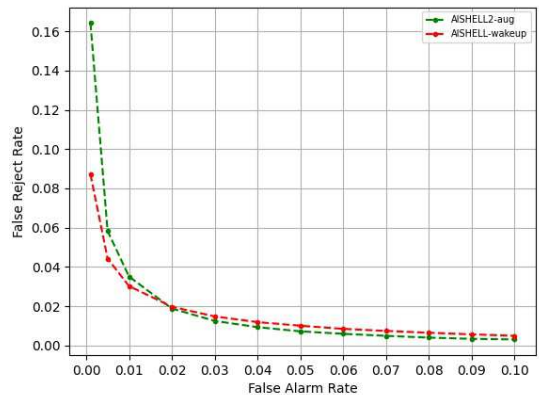}}
\caption{ROC curves when speeches in AISHELL-wakeup or AISHELL2-aug are chosen as positive samples}
\label{fig:wake-up-roc}
\end{figure}

A deep-neural network(DNN) based keyword spotting system is firstly tested on AISHELL-wakeup, AISHELL2-aug and AISHELL2 (speeches in AISHELL-wakeup or AISHELL2-aug are positive samples, speeches in AISHELL2 are negative samples). The DNN with 7 hidden layers and 256 nodes per hidden layer is pre-trained with about 5000h speeches. The DNN has 411 output labels: 409 Chinese characters, a silence label and a Filler label for music and noise. 40-dimensional fbank features are extracted with a frame shift of 20ms and a window width of 30ms and then 5 future frames and 5 frames in the past are stacked to predict posterior probabilities for each output label using the DNN model. The posterior handling module proposed in~\cite{2014Small} combines the label posteriors produced every frame into a confidence score used for detection. Figure~\ref{fig:wake-up-roc} shows the performance when speeches in AISHELL-wakeup or AISHELL2-aug are chosen as positive samples. Results are demonstrated in the form of Receiver Operating Characteristic (ROC) curves, where the false reject rate(that is, a key phrase is present but a negative decision is given, FRR) is on the Y-axis and the false alarm rate (that is, a key-phrase is not present, but a positive decision is made, FAR) is on X-axis. The ROC is obtained by sweeping through confidence thresholds.

As is shown in Figure~\ref{fig:wake-up-roc}, when FAR is larger than 0.01, speeches in AISHELL2-aug could achieve very similar FRR with speeches in AISHELL-wakeup. This means the synthesized speech does contain a phonetic context of "ni"-"hao"-"mi"-"ya". As the synthesised speeches are not natural-sounding, when FAR drops to 0.001, FRR of AISHELL-aug is 0.164 while that of AISHELL-wakeup is 0.087. In our future work, we will try to produce or select more natural-sounding synthesized speeches to train the x-vector model.

\begin{figure}[!htb]
\centering
\includegraphics[width=80mm]{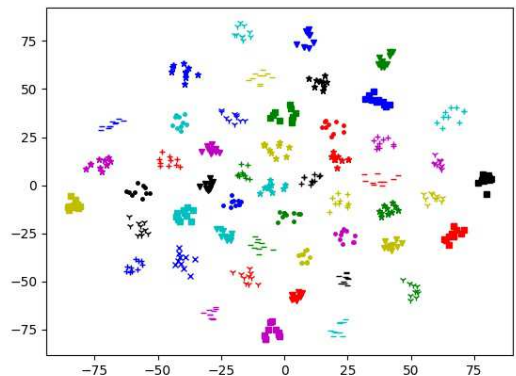}
\caption{t-SNE visualization of embeddings from 50 random speakers of AISHELL2-aug, samples with the same shape and color are from the same speaker}
\label{fig:t-SNE}
\end{figure}

The text-independent x-vector model trained with the Voxceleb2 database in our previous work~\cite{xiang2019margin} is used to extract embeddings from audios of AISHELL2-aug. 50 speakers are randomly chosen, and their embeddings are shown in Figure~\ref{fig:t-SNE} using t-SNE. Figure~\ref{fig:t-SNE} shows the good speaker discriminative property of synthesized speeches.

\section{Conclusions and future work}
In this work, we proposed a unit selection synthesis based data augmentation for text-dependent speaker verification. The proposed method enables us to leverage the rich text-independent data to fast generate new speech with the desired transcript, which leads to a better-performing and more robust text-dependent speaker verification system. This strategy can reduce the development period and cost dramatically. Experiments on the AISHELL-2019B-eval corpus shows that the proposed approach could achieve a relative performance improvement of about 40\% in both EER and minDCF.

In the future work, we will focus on synthesizing more natural-sounding and variable speech to further increase the robustness of speaker verification systems.

\bibliographystyle{IEEEbib}
\bibliography{strings,refs}

\end{document}